\documentclass[%
 reprint,
 amsmath,amssymb,
 aps,
]{revtex4-1}
\UseRawInputEncoding
\usepackage{graphicx}
\usepackage{dcolumn}
\usepackage{bm}
\usepackage{natbib}

\begin{document}
\preprint{APS/123-QED}

\title{Experimental observation of topological transition in linear and non-linear parametric oscillators}
\author{Benjamin Apffel}
 \email{benjamin.apffel@epfl.ch}
\affiliation{%
Institute of Electrical and Micro Engineering, Laboratory of Wave Engineering, Ecole Polytechnique Fédérale de Lausanne (EPFL), Station 11, 1015 Lausanne, Switzerland
}%
\author{Romain Fleury}%
\affiliation{%
Institute of Electrical and Micro Engineering, Laboratory of Wave Engineering, Ecole Polytechnique Fédérale de Lausanne (EPFL), Station 11, 1015 Lausanne, Switzerland
}%

\date{\today}

\begin{abstract}
Parametric oscillators are examples of externally driven systems that can exhibit two stable states with opposite phase depending on the initial conditions. In this work, we propose to study what happens when the external forcing is perturbed by a continuously parametrized defect. Initially in one of its stable state, the oscillator will be perturbed by the defect and finally reach another stable  state, which can be its initial one or the other one. For some critical value of the defect parameter, the final state changes abruptly. We  investigate theoretically and experimentally such transition both in the linear and non-linear case, and the effect of non-linearities is discussed. A topological interpretation in terms of winding number is proposed, and we show that winding changes correspond to singularities in the temporal dynamics. An experimental observation of such transition is performed using parametric Faraday instability at the surface of a vibrated fluid. 
\end{abstract}

\maketitle

\section{Introduction}
The parametric oscillator is one of the simplest externally driven system, a paradigmatic example being a pendulum shaken vertically at at twice its resonance frequency.  Parametric oscillators are encountered in many different fields of physics including quantum mechanics \cite{burnham_observation_1970,milburn_production_1981,wu_squeezed_1987}, optics \cite{akhmanov_observation_1965,giordmaine_tunable_1965}, (micro-)mechanics \cite{grandi_enhancing_2021,carr_parametric_2000,rugar_mechanical_1991,turner_five_1998}, electronics \cite{berthet_analog_2002} or hydrodynamics \cite{faraday_xvii_1997,douady_experimental_1990}, making their study of prime interest. The rich physics of this system results from complex interactions between damping, forcing and non-linearity. The response amplitude of the system strongly depends on the exact parameters of the system and can exhibit (among others) linear stability around zero, sub-critical or super-critical Hopf bifurcation toward a finite amplitude oscillation, sub-harmonic oscillation, bistability and hysteresis \cite{godreche_pattern_1998,douady_experimental_1990}. For a given configuration, the amplitude response when the external forcing is perturbed has also been investigated recently in the context of time crystals \cite{dhardemare_probing_2020,lustig_topological_2018}.

In parallel to the amplitude behavior, the phase response of the system has also been investigated in several regimes. In the linearly stable regime, it can for instance be used to perform squeezing of noise in thermal \cite{rugar_mechanical_1991} and quantum \cite{milburn_production_1981,wu_squeezed_1987} regimes. In the sub-harmonic finite amplitude oscillation regime, the oscillation phase is fixed by the forcing and can only take two discrete values separated by $\pi$. This phase degeneracy has for instance been used as an analog bit to store information \cite{goto_parametron_1959,sterzer_microwave_1959}, or to simulate the two states of a $1/2$-spin in coherent Ising machines to compute the ground state of Hamiltonians \cite{utsunomiya_mapping_2011,honjo_100000-spin_2021,inagaki_large-scale_2016, wang_coherent_2013, mcmahon_fully_2016}. In this context, several strategies to perform bit-flipping (or phase switch) have been implemented by perturbing the excitation signal in a specific manner \cite{mahboob_bit_2008,frimmer_rapid_2019}. Possible bifurcation between several stables states in the presence of noise has also been investigated in the context of chaos control \cite{luchinsky_optimal_2002,boccaletti_control_2000, beri_solution_2005}. 

The purpose of the present work is to study the condition for a phase switch to occur by considering a continuously parametrized family of perturbations (or time defects) rather than random noise or specific perturbations. The continuous parameter will generate a family of trajectories which will display strong discontinuities at specific values of the parameter. The paper is organized as follow. We first introduce some useful formalism to study linear parametric oscillators and define our family of time defects. We then show that their exists some critical values of the parameter that separate no phase switch to phase switch behavior, giving rise to a transition with respect to the defect parameter. Interestingly, we show that this transition exhibits topological properties and provide experimental observation of such transition. We then introduce some non-linearity and discuss its impact theoretically, numerically and experimentally on the transition process. 
\section{Results}
\subsection{Parametric oscillation and averaging equation}
\begin{figure}
    \centering
    \includegraphics[width=8.6cm]{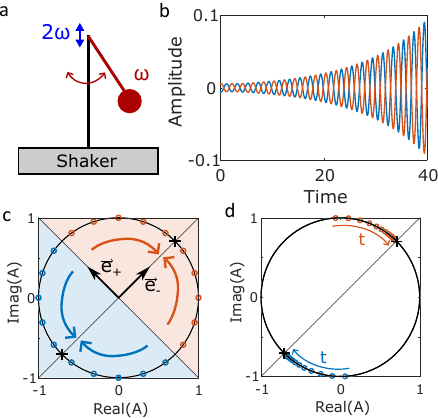}
    \caption{\textbf{Parametric oscillator and phase state.} (a) An example of a parametric oscillator: a pendulum with natural pulsation of $\omega$ vibrated at $2\omega$. (b) Angular position of the pendulum evolving according to Eq. \ref{eq:PO} for $\omega= \pi$, $\epsilon=0.03$, $\omega \tau=200$ and opposite initial conditions.Depending on the later, two phases of the oscillation can be observed. (c) Diagram showing for different initial conditions on the unit circle which phase state is reached by the system at long time. The two regions are defined by $\vec A_0. \vec e_- > 0$ (red) and $\vec A_0. \vec e_- < 0$ (blue). (d) Normalized complex amplitude along time computed from (b). The oscillation phase converges toward one of the two fixed points as time flows. }
    \label{fig:1}
\end{figure}
 We first describe here the key features of parametric oscillation and phase selection. We consider as in Fig. \ref{fig:1}a a pendulum with natural pulsation $\omega$ and vibrated at $2\omega$. For now we will not consider non-linear effects and the evolution equation for the angular coordinate $\psi$ is then
\begin{equation}
\frac{d^2 \psi}{dt^2} + \frac{2}{\tau} \frac{d \psi}{dt} +  \omega^2(1 + \epsilon \cos{(2 \omega  t)} ) \psi = 0
\label{eq:PO}
\end{equation}
where $\tau$ is the damping time and $\epsilon$ is the adimensioned forcing amplitude. We will assume here that the forcing, the damping and the initial condition are small: $\epsilon \ll 1, 1/\omega \tau = \epsilon \gamma \ll 1$, $\psi \ll 1$, $ \dot \psi = 0$. For $\gamma <1/2$, the angular position $\psi(t)$ exponentially grows with a slow time scale while oscillating at $\omega$ as shown in direct numerical simulation of Eq. \ref{eq:PO} in Fig. \ref{fig:1}a. In order to discard the oscillation, we introduce the complex envelope $A(t)$ as
\begin{equation}
\psi(t) = Re(A(t) e^{i\omega t})  \qquad \dot \psi(t) = - \omega Im(A(t) e^{i\omega t}) 
\label{eq;VdPtrans}
\end{equation}
This envelope will evolve with a slow time scale $\sim \epsilon/\omega$ (see Methods or \cite{papariello_parametric_2018}) and carry information on the oscillation phase. If one identifies the complex amplitude $A(t)$ with its corresponding vector $\vec A$ in $\mathbb{R}^2$, it evolves according to
\begin{equation}
    \partial_t \vec A = \frac{\partial}{\partial t}\begin{bmatrix}
        Re(A) \\ Im(A)
    \end{bmatrix}
    = -\epsilon
    \begin{bmatrix}
        \gamma/2 & -1/4 \\
        -1/4 & \gamma/2 \\
    \end{bmatrix}
    \begin{bmatrix}
        Re(A) \\ Im(A)
    \end{bmatrix}
\end{equation}
The matrix admits two eigenvalues $\lambda_\pm = - \epsilon( \gamma/2 \pm 1/4)$ with corresponding orthogonal eigenvectors $\vec e_{\pm} = [1;\mp 1]^t$. If the damping is small enough ($\gamma < 1/2$), one has $\lambda_+ < 1 < \lambda_-$. For the angular coordinate $\psi(t)$, three behavior can therefore be observed when $t \rightarrow \infty$ depending on the initial condition $\vec A_0$: 
\begin{itemize}
    \item if $\vec A_0.\vec e_- > 0$, $\psi(t)$ exponentially grows while oscillating at $\omega$ with phase $\pi /4$ 
    \item if $\vec A_0.\vec e_- < 0$, $\psi(t)$ exponentially grows while oscillating at $\omega$ with phase $\pi /4 + \pi$ 
    \item if $\vec A_0.\vec e_- = 0$, $\psi(t)$ goes exponentially to zero
\end{itemize}
Unless specific initial conditions are chosen and the system is completely free of noise, the last behavior is not seen and the generic behavior is therefore an exponentially grows combined with an oscillation which can only have two different phases. This 2-fold phase degeneracy is directly related to the 2:1 resonance between the forcing and the pendulum response combined with the invariance of the evolution equation under the discrete symmetry $t \rightarrow t + \pi/\omega$. One can also understand it geometrically : as the projection of $\vec A_0$ on $\vec e_+$ will be exponentially killed, $\vec A$ will tend to align or anti-align with $\vec e_-$. The choice of one phase or the other is therefore determined by the sign of the projection between the initial condition and the growing eigenvector $\vec A_0.\vec e_-$ as shown in Fig. \ref{fig:1}c. The degeneracy is illustrated in Fig. \ref{fig:1}b where we perform a direct simulation of Eq. \ref{eq:PO} with the same forcing but opposite initial conditions. As everything is linear, it is enough to consider the normalized amplitude $\Tilde{A} = A(t) / |A(t)|$. Its evolution is shown in Fig. \ref{fig:1}d and its convergence toward one of the two 'phase states' is visible.

\subsection{Introducing a continuously parameterized perturbation}

\begin{figure*}
    \centering
    \includegraphics[width=18cm]{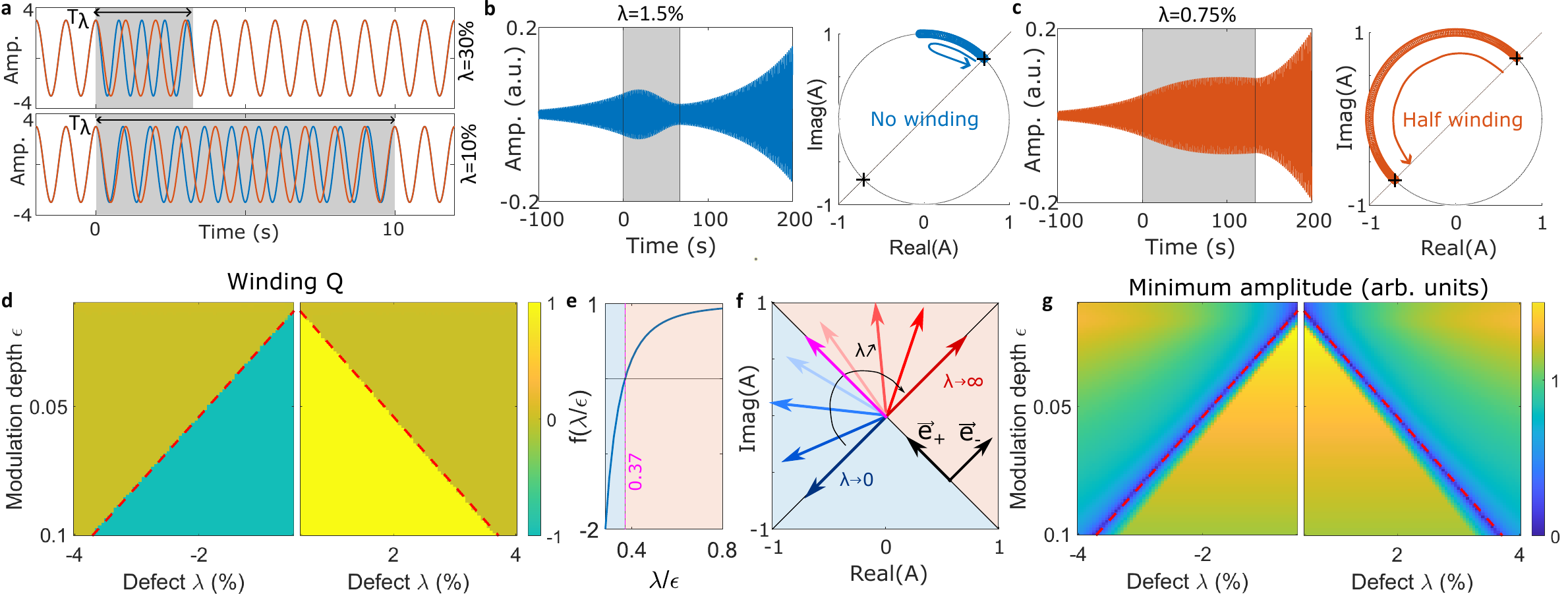}
    \caption{\textbf{Temporal defect, phase state change and topological aspect of transition in the linear case.} (a) Two examples of defects (blue curves) for $\lambda= 30 \%$ (top) and $\lambda = 30 \%$ (bottom), the unperturbed excitation being plotted for comparison in red. The duration of the detuning is chosen so that the two excitation signals collapse back together after $T_\lambda$. (b) Simulated evolution of the parametric oscillator for $\epsilon = 0.03$, $\lambda = 1.5 \%$ and (c) $\lambda = 0.75 \%$, the grey area corresponding to the defect duration. The phase of the corresponding complex amplitude is shown besides, and the change of phase state only occurs for the second case. (d) Half-winding $Q$ of the normalized amplitude $\Tilde{A}(t)$ for different values of ($\epsilon, \lambda$). The red dashed line corresponds to the analytical prediction $\lambda/\epsilon = 0.37$.  (e) Analytical computation of $\vec A(T_\lambda).\vec e_- = f(\lambda/\epsilon$ which cancels for $\epsilon/\lambda=0.37$. (f) Normalized complex amplitude $\Tilde{A} (T_\lambda)$ when $\lambda$ varies from $0$ to infinity, the initial amplitude being $\vec A(0) = \vec e_-$. The transition from positive to negative (pink arrow) occurs for $\lambda/\epsilon = 0.37$.  (g) Minimum amplitude of the field min$_{t>0}|A(t)|$ for different values of  $(\lambda, \epsilon)$. The field cancels exactly when the winding changes.}
    \label{fig:2}
\end{figure*}

Let us now assume that the forcing amplitude is fixed and that the parametric oscillator has reached one of its phase state. We now perturb the forcing of the system by detuning between $t=0$ and $t = T_{\lambda} = \pi / \omega |\lambda|$ the excitation frequency from $2 \omega$ to $2\omega (1 + \lambda)$, all other parameters being left unchanged. Doing so, the perturbed excitation takes an overall phase of $\pm 2 \pi$ (depending on the sign of the detuning $\lambda$) compared to the unperturbed signal, as shown in Fig. 2a. After $T_\lambda$, the perturbed excitation is back in phase with the unperturbed one and the excitation frequency is set back to $2 \omega$. This particular choice of $T_\lambda$ ensures (1) the continuity of the excitation signal as shown in Fig. \ref{fig:2}a and (2) that the possible phase states of the oscillator are the same before and after the defect, since they are completely determined by the phase of the forcing. Equivalently, this condition ensures that the eigenvectors for the envelope equation are the same before and after the defect.

\subsection{Discrete phase states vs continuous parameter}
Adding this defect will perturb the parametric oscillator initially in one of its phase state (aligned or anti-aligned with $\vec e_-$). As everything is linear, let us assume without loss of generality that $A(t=0) = \vec e_-$. During the defect, the evolution equation is modified so that $\vec e_\pm$ are no more eigenvectors for the evolution operator (see Methodes). The exponential growth will be perturbed and $\vec A$ will start to move away from $\vec e_-$ as shown in Fig. \ref{fig:2}b-c. After the defect, the system is in the state $\vec A(T_\lambda)$ and will after some time align or anti-align again with $\vec e_-$. As discussed before, knowing which of the two occurs is fully determined by the sign of the geometrical projection $\vec A(T_\lambda).\vec e_-$. The introduction of the defect can therefore be seen as a change in the initial condition, and varying the defect parameter $\lambda$ as a continuous change of the initial conditions. 

Interestingly, both phase change and no phase change can occur depending on the value of $\lambda$. This is shown in Fig. \ref{fig:2} b-c where for the same forcing $\epsilon = 0.03$, a phase change occurs for $\lambda =0.75 \%$ but not for $\lambda = 1.5 \%$. In order to characterize the phase difference between the initial and the final state, we define the number 
\begin{equation}
    Q(\lambda, \epsilon) = \frac{1}{\pi}\Im{\left( \log{\frac{A(\infty)}{A(0)} }\right)}
    \label{eq:Ai/A}
\end{equation}
that counts the number of half-winds around the origin, which must be an integer. The resulting transition diagram in the $(\lambda, \epsilon)$ plane shows whether a phase change occurs or not depending on the system parameters. Direct numerical computation of the half-winding number $Q(\lambda, \epsilon)$ is displayed in Fig. \ref{fig:2}d. It is constant on large domains that are separated by a curve of critical values of $\lambda$ that can be fitted as $ \lambda_c = \alpha \epsilon$ with $\alpha \approx 0.37$ (red line).

 As discussed above, the transition is entirely driven by the sign of $\vec A(T_\lambda).\vec e_-$. An explicit computation using perturbative methods shows that if $\vec A(t=0) = \vec e_-$, one has $\vec A(T_\lambda).\vec e_-= f(\lambda/\epsilon)$ where $f$ is plotted in figure \ref{fig:2}d and given explicitly in Methods section. In particular, it only cancels once for $\lambda/ \epsilon \approx 0.37$, which exactly matches direct numerical simulation (see Fig. \ref{fig:2}e). Such analysis provides moreover a straightforward geometrical interpretation of the transition. For a given forcing $\epsilon$, the vector state $\vec A(T_\lambda)$ performs half a turn between $\vec e_-$ and $-\vec e_-$ when $\lambda$ goes from 0 to $\infty$ as illustrated in Fig. \ref{fig:2}f. The transition between no phase shift and phase shift occurs at the critical value $\lambda_c$ when  the vector state becomes exactly orthogonal to $\vec e_-$. 

At this critical value, the projection cancels out and $\vec A(T_\lambda)$ is aligned with $\vec e_+$. Accordingly, the parametric oscillator is expected to go exponentially fast to zero. This is confirmed by the diagram in Fig. \ref{fig:2}g that shows the (normalized) minimum amplitude reached by the oscillator $\min_{t>0} |A(t)|$. The amplitude therefore exactly cancels out when the winding change occurs.

\subsection{Topological aspect of the transition}
Such cancellation near winding number change is reminiscent of what typically occurs in condensed matter during a topological phase transition. In this context, topological invariants are written as an integral over the reciprocal space \cite{batra_physics_2020,kitagawa_topological_2010,fleury_floquet_2016}. If one writes $A(t) = e^{i \phi(t)}$, one can also write $Q$ as an integral but it will rather run over the time dimension
\begin{equation}
    Q(\lambda)= \frac{1}{\pi} \int_0^\infty \dot \phi dt
    \label{eq:Qint}
\end{equation}
In terms of topology, the winding number $Q$ characterizes the homotopy class of the path followed by the complex amplitude. The variation of $\lambda$ can therefore be seen as the generation of a family of continuously parameterized trajectories in the complex plane. For particular values of $\lambda$, the homotopy class of the trajectory suddenly changes corresponding to a discontinuity of path parametrization with respect to $\lambda$. This occurs when the field reaches the origin and cancels out. Topological transition are usually associated to singularities. In the present case, it is the singular behavior of the dynamics at this specific value of $\lambda$ that plays this role. Indeed, the origin is an unstable equilibrium point, and it can therefore only be reached for very specific values of $\lambda$ while the generic behavior is to reach one of the two stable state (aligned an anti-aligned with $\vec e_-$).

One can push the analogy with topological transitions a bit further. A phase transition in condensed matter is typically associated with the divergence of some correlation length at the transition. In our case, such length is replaced by the typical convergence time $T_c$ taken by $\vec A(t)$ to approximately align again with $\vec e_-$. Close from the critical value $|\lambda-\lambda_c| \ll 1$, one expects $T_c \sim \log |\lambda- \lambda_c|$. This scaling has been verified numerically (see section F), and one therefore finds an exponential scaling law rather than a power law with some critical exponents in condensed matter.

\subsection{Experimental observation}
\begin{figure}
    \centering
    \includegraphics[width=8.5cm]{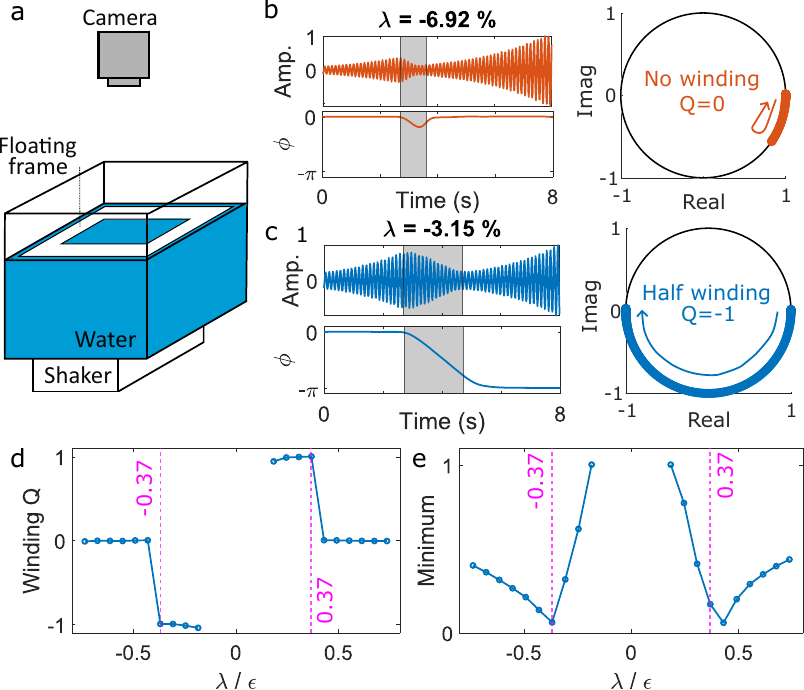}
    \caption{\textbf{Experimental measurement of phase state change in linear regime.} (a) Sketch of the experimental setup: a tank filled with water is placed on a shaker vibrating at 15.92 Hz. A floating frame at the surface ensures a good definition of the first parametrically unstable mode at 7.92 Hz. A camera on the top recovers the wave amplitude through demodulation of a periodic pattern placed below the surface (not shown here). (b) Typical experimental signal for the wave amplitude (in arbitrary units) and the slowly varying phase for $\lambda = -6.92 \%$ and (c) $\lambda= -3.15 \%$. In both cases, the forcing is $\epsilon=0.10$ and the defect occurs at $t\approx 4 s$. Half negative winding occurs in the second case but not in the first one. (d) Half-winding number $Q(\lambda)$ and (e) minimum amplitude for different values of $\lambda$ plotted as a function of $\lambda/\epsilon$. The vertical lines show the theoretical prediction for the transition  $\lambda_c/\epsilon=0.37$.}
    \label{fig:3}
\end{figure}
We now propose an experimental observation of such transition between different homotopy classes. As a parametric oscillator, we chose the Faraday waves that occurs at the surface of a vibrated fluid \cite{douady_experimental_1990}. The experimental setup consists in a tank of water vibrated at $2\omega$ using a shaker. In order to avoid friction of the parametrically excited waves on the walls, a plastic frame pierced in its center by a square hole of 2 $\times 2$ cm is placed at the surface as shown in Fig. \ref{fig:3}a. The frame is floating on the water so that evaporation of water does not change the boundary condition of the meniscus, which could impact the experiments \cite{douady_experimental_1990, dhardemare_probing_2020}. The first mode of the square hole is measured to occur at $7.92 \pm 0.02$ Hz (see Methods) and we therefore fix the excitation frequency to $15.84$ Hz. The instability threshold is found for a critical acceleration of $a_c \approx 1.4$ m/s$^2$ corresponding to $\epsilon \approx 0.11$ (see Methods). In this case, we observe waves at half the excitation frequency that exponentially grows for typically 15 s before saturating at finite amplitude due to non-linearity. As we aim to probe the linear regime, each experiment consists in 7 s of unperturbed excitation, a defect parametrized by $\lambda$ and finally 8 s of unperturbed excitation again before stopping the vibration and let the system come back to rest for the next measurement. Such procedure ensures that the wave amplitude remains far from its non-linear saturation. The wave amplitude is recovered by a camera placed above the water and a periodic pattern placed below the surface. Using demodulation algorithm, one can retrieve the surface deformation through the pattern distortion recorded by the camera \cite{wildeman_2017}.

\begin{figure*}
    \centering
    \includegraphics[width=17cm]{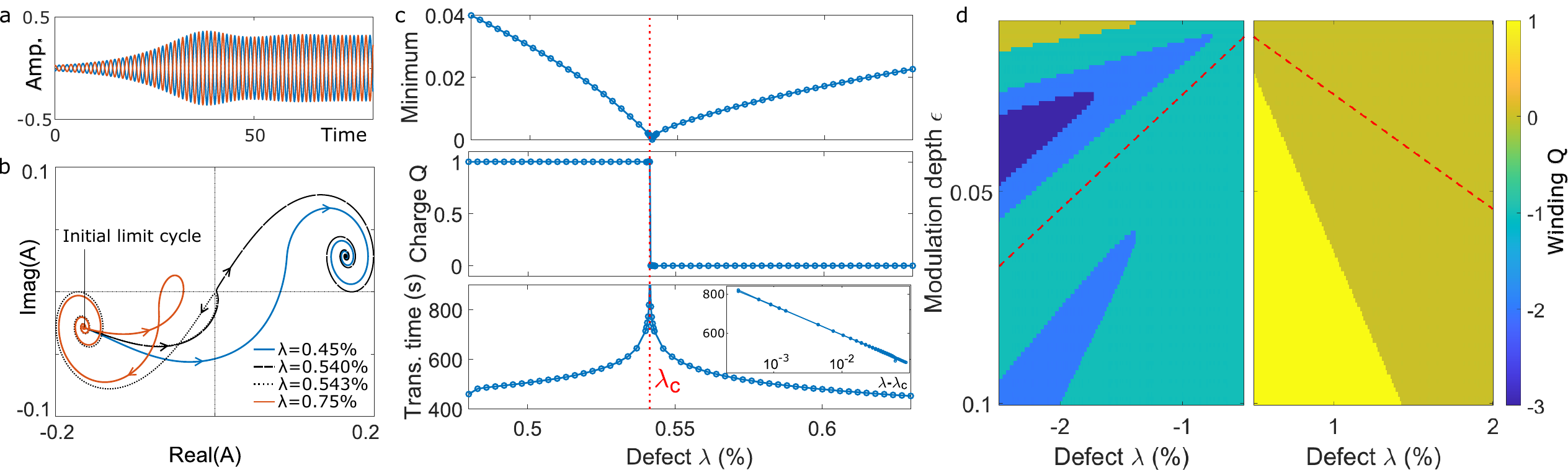}
    \caption{\textbf{Impact of non-linearity on the transition process.} (a) Angular position along time simulated from non-linear evolution with $\omega=\pi, \epsilon=0.03, \omega \tau=200$ for opposite initial conditions. Depending on the initial condition, two phases for the oscillation can be observed. Contrary to the linear case, the system reaches finite amplitude oscillation (limit cycles). (b) Trajectories in the complex plane for $\epsilon=0.03$ and $\lambda=0.45, 0.540, 0.543, 0.6 \%$ when the system is intially in one of its limit cycles. The two convergence points of the trajectories are the two possible limit cycles with opposite phase. (b) Half-winding index $Q$ (up), minimum amplitude along time min$_{t>0}|A(t)|$(middle) and time of convergence $T_c$ (bottom) as a function of $\lambda$. Inset: same plot with logarithmic horizontal axis to test $T_c \sim \log|\lambda-\lambda_c|$ (c) Transition diagram $Q(\epsilon, \lambda)$ for the non linear case. The red-dashed line corresponds to the prediction from the linear case.}
    \label{fig:4}
\end{figure*}

 Some typical wave amplitude measurement along time are shown in Fig. \ref{fig:3}b-c with $\lambda = -6.92 \%$ and $\lambda= -3.15 \%$ respectively. From these curves, one can compute the phase and the corresponding winding in the complex plane as shown in the same panels. Depending on the value of $\lambda$, the trajectory either performs  half negative, half positive or no winding around the origin of the complex plane. The Fig. \ref{fig:3}d shows the value of the winding index $Q$ for different values of $\lambda$, the forcing being fixed to $\epsilon=0.10$. In order to compare our results with the theory, we normalize the horizontal axis by $\epsilon$. The resulting curve is antisymmetric with respect to $\lambda/\epsilon$, as expected from theory and numerical simulations. Moreover, the transition occurs for a specific value of $\lambda/\epsilon$ that is consistent with the theoretical prediction $\lambda_c/\epsilon = \pm 0.37$. We also compute the minimum amplitude of the waves normalized by the maximal amplitude before the defect and the result is shown in Fig. \ref{fig:3}e. The transition between different winding numbers indeed corresponds to an (almost) cancellation of the wave amplitude, which is again consistent with the prediction.

It is tempting to vary $\epsilon$ in order to establish complete transition diagrams. However, as the later increases, the system hits non-linearities within a few seconds, preventing to perform measurement in the purely linear case. The next sections are dedicated to discuss the role of non-linearities theoretically and experimentally in the transition process.

\subsection{Adding non-linearity}
Realistically, all parametric oscillators will not grow exponentially forever but will eventually saturate in amplitude due to non-linearity. We therefore propose here study how the later can impact the previous analysis. For this, we take a simple model and introduce a cubic term in Eq. \ref{eq:PO} such that the evolution equation is now
\begin{equation}
\frac{d^2 \psi}{dt^2} + \frac{2}{\tau} \frac{d \psi}{dt} +  \omega^2(1 + \epsilon \cos{(2 \omega  t)} ) (\psi-\alpha \psi^3) = 0
\label{eq:PONL}
\end{equation}
The non-linearity tends to saturate the oscillation amplitude and Eq. (\ref{eq:PONL}) admits two sub-harmonic oscillating limit cycles with phase difference of $\pi$ as shown in Fig. \ref{fig:4}a. Depending on the initial conditions, one or the other will be chosen in a similar fashion as in the linear case. The limit cycles are associated with constant complex amplitude $A(t)$ and can take two values $\pm A_c$ that correspond to the two possible phase state of the system. Convergence toward those fixed points in the complex plane replace alignment of $\vec A$ with $\pm e_-$ in the linear case. As the problem is no longer linear, we will moreover consider the trajectories of $A(t)$ in the complex plane rather than only its phase. 

We assume that the system ran for a long time before $t=0$ and that it reached a limit cycle. Adding a defect parameterized by $\lambda$ will perturb the cycle, and the complex amplitude will start to move in the complex plane. After the perturbation, the system reaches one of its two limit cycles. Numerical simulations for different values of $\lambda$ are are shown in Fig. \ref{fig:4}b, the forcing being fixed to $\epsilon=0.03$. Depending on the value of $\lambda$, a phase change occurs or not, which is associated respectively with non-zero or zero half winding around the origin. When $\lambda$ spans different values, a family of trajectory is generated in the same manner as in the linear case.

The half-winding index $Q$ is shown in Fig. \ref{fig:4}c (upper panel) for different values of $\lambda$  and goes abruptly from one value to the other for specific value $\lambda_c \approx 0.54$ \%. In terms of path, it is shown in Fig. \ref{fig:4}b that close values of $\lambda =0.540 \%$ and $\lambda =0.543 \%$  can correspond to very different trajectories. As in the linear case, the transition is associated with a zero amplitude (middle panel). One can also compute the time for the system to converge back to its new limit cycle, and this time diverges at the transition (lower panel). Moreover, the scaling law $T_c \sim \log|\lambda-\lambda_c|$ predicted in the linear case still holds as shown in the inset with a semi logarithmic plot. For $\lambda \sim \lambda_c$ the amplitude is very small during most of the transition between the limit cycles. The non-linearity can therefore be neglected during most of the dynamics, leaving the scaling law for the transition time insensitive to it.

Many works have investigated the switching between stable states bifurcation in the presence of noise \cite{beri_solution_2005,silchenko_fluctuational_2003}. It has been shown in particular that the system tends to escape through heteroclinic trajectories, which is exactly the one observed at critical values of $\lambda$. This heteroclinic trajectory plays the role of the eigenvector $\vec e_+$ in the linear case and separates different homotopy class for the trajectories.  We would also like to mention that changing the type of defect would change the continuous family of trajectories generated but not the physical features of transitions (cancellation of field, divergence of the convergence time...). The framework presented here may therefore apply to other situations with different types of defects.

\subsection{Impact of non-linearity on transition diagrams}
For now, it seems that apart from replacing eigenvectors by limit cycles to take into account amplitude saturation, non-linearity may not significantly impact the physics of transition. This is actually not the case. Surprising consequences of non-linearity are observed when one computes the transition diagram $Q(\epsilon, \lambda)$ shown in Fig. \ref{fig:4}d, that is completely different from the one shown in Fig. \ref{fig:2}d. First, it clearly breaks the symmetry between positive and negative values of detuning $\lambda$. Second, larger winding number are observed in those diagrams for negative detuning but not for positive ones. It has also been verified that each transition of winding number is associated with field cancellation and convergence time divergence as in the linear case. Nevertheless, the values of $(\lambda, \epsilon)$ at which the transition occurs are completely different from the linear case as shown in Fig. \ref{fig:4}d.  

We briefly discuss here the reasons for such differences betwenn linear and non-linear regimes. One can show that the evolution equation for the amplitude is of the form
\begin{equation}
    \dot A + \frac{1}{\omega \tau} A=  - \frac{3 i \alpha \epsilon}{2} |A|^2 A + \frac{i \epsilon}{4} A^* e^{2 i  \lambda(t) t}
    \label{eq:aNL}
\end{equation}
where $\lambda(t) = \lambda$ for $0 \leq t \leq T_\lambda$ and zero otherwise. In the linear limit $\alpha = 0$, one can map this equation on another one that only depends on $\lambda/\epsilon$ (see Methods). This indicates that the results in the linear case can only depend on this ratio, which is consistent with the results from the previous section. A similar argument predicts the antisymmetry $\lambda \rightarrow - \lambda$ in the linear case. However, both of these symmetries break down in the presence of non-linearity ($\alpha \neq 0$). This is reminiscent of the different behaviors of parametric oscillators (subcritical or supercritical bifurcation) depending on the sign of the detuning \cite{douady_experimental_1990,godreche_pattern_1998}. Such analysis explains why the phase diagram are different and non-symmetric in the non-linear case, but do not predict when the transition occurs. Such prediction requires a full treatment of the non-linear problem, which appears very difficult to perform.

\subsection{Experimental observation}

\begin{figure}
    \centering
    \includegraphics[width=8cm]{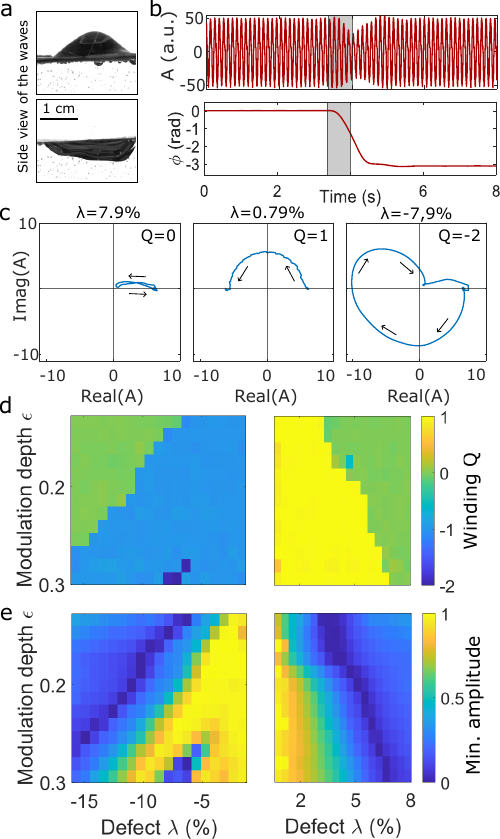}
    \caption{\textbf{Experimental measurement of transition diagram in the non-linear case.} (a) Pictures from the side of the parametrically excited water wave limit cycles with opposite phases. (b) Typical experimental measurement of wave amplitude along time and its associated complex phase. The grey area materializes the time defect. (c-e) Complex amplitudes showing different half-winding values for fixed $\epsilon=0.29$ : $Q=0$ (right, $\lambda=7.9 \%$), $Q=1$ (middle, $\lambda=0.79 \%$), $Q=-2$ (left, $\lambda=-7.9 \%$) (d) Experimental transition diagram of $Q$ and (e) associated minimum amplitude of the wave field.}
    \label{fig:5}
\end{figure}
Last, we propose an experimental observation of a non-linear transition diagram. We use the same experimental setup as before, for which the instability occurs for $a \sim 1.3$ m/s$^2$ which corresponds to $\epsilon_c \sim 0.12$ (see Methods). On the other hand, taking $a$ typically greater than $3$ m/s$^2$ leads to unwanted phenomena such as water droplet ejection, limiting the maximal reachable forcing to $\epsilon \sim 0.3$. We perform wave amplitude measurement for various values of $\epsilon$ and $\lambda$ in order to obtain the complete transition diagram. Contrary to the linear case, we let the system reach its saturated oscillation amplitude before introducing the temporal defect. This occurs within a few seconds for $\epsilon > 0.13$. Due to the high amplitude of the non-linear waves, the previous demodulation method used to recover the wave amplitude could not be used. We therefore place a camera on the side and recover the wave amplitude using contrast analysis from raw pictures as in Fig. \ref{fig:5}a. Typical measurement of the wave amplitude along time as well as the corresponding phase are shown in Fig. \ref{fig:5}b (see Methods for details). Several examples of complex amplitude $A(t)$ with different half-winding numbers associated to different values if $\lambda$ are also shown in Fig. \ref{fig:5}c.

We performed systematic measurement of half-winding $Q$ for different values of forcing $\epsilon$ and detuning $\lambda$. The obtained transition diagram is shown in Fig. \ref{fig:5}d. It clearly exhibits distinct domains with different values of the winding index $Q$. We see once again the existence of large domains separated by critical curves. At those critical curves, the minimum amplitude of the field vanishes as shown in Fig. \ref{fig:5}e, which is consistent with the previous analysis. However, there exists a strong asymmetry between positive and negative values of $\lambda$ which was not seen in the linear regime. Moreover, larger winding index of -2 have also been observed experimentally as shown in Fig \ref{fig:5}c-d. Although quite simple, the non-linear model discussed above exhibits similar features. The domains in Fig. \ref{fig:4}d and \ref{fig:5}d look moreover similar qualitatively, but could not be matched quantitatively. This suggests that the full dynamic of Faraday waves is not entirely captured by Eq. \ref{eq:aNL}, which was derived for small forcing.

  \section{Conclusion}
In this article, we have studied theoretically and experimentally how a system that exhibits a discrete number of limit cycles can switch (or not) from one to another by introducing continuously parameterized defects in the excitation. Both the linear and non-linear case have been discussed, and the difference between the two have been emphasized. Our results provide an example of a topological transition in a non-linear time varying system, and the topological index can be interpreted as a half-winding number of trajectories in the phase space. All the predicted features of the transition have been observed experimentally using Faraday instability. The results presented in this article may find applications in control of parametric oscillators or for robust bit flip in coherent Ising machines. It also paves the way toward the definition of new topological features in non-linear or active systems that exhibit finite number of limit cycles.

\section{Methods}

\subsection{Perturbative treatment of a parametric oscillator}

We consider a parametric pendulum with natural frequency $\omega_0$, damping time $\tau$ driven with adimensioned acceleration $\epsilon$ oscillating at $\omega$ and a cubic non-linearity.
\begin{equation}
\frac{d^2 \psi}{dt^2} + \frac{2}{\tau} \frac{d \psi}{dt} +  \omega_0^2(1 + \epsilon \cos{(2\omega  t)} ) (\psi - \alpha \psi^3 ) = 0
\end{equation}
For the case of a hanging pendulum, one has $\alpha =1/6$ to recover the Taylor expansion of $\sin(\psi)$. In all that follows, we will assume that the forcing is small $\epsilon \ll 1$ and expand the solutions with respect to this parameter. We aim to show here that this equation admits two harmonic limit cycles with a difference phase of $\pi$ when the forcing is small.

We will also assume that the viscous term scales as $2/\tau = \gamma \omega_0 \epsilon$ and rescale $\psi = \sqrt{\epsilon}{\psi}$. The equation can now be written as
\begin{equation}
\frac{d^2 \psi}{dt^2} +  \gamma \omega_0 \epsilon \frac{d \psi}{dt} +  \omega_0^2 (1 + \epsilon \cos{(2\omega  t)} ) (\psi - \epsilon \alpha \psi^3 ) = 0
\end{equation}
Last, we set a new time $t \rightarrow \omega_0 t$ so that the equation finally reads

\begin{equation}
\frac{d^2 \psi}{dt^2} +  \gamma \epsilon \frac{d \psi}{dt} + (1 + \epsilon \cos{(2 \Omega t)} ) (\psi - \epsilon \alpha \psi^3 ) = 0
\end{equation}
with $\Omega = \omega/\omega_0$. We will now establish an envelope equation using averaging method (see for instance \cite{papariello_parametric_2018} pp.8-12 and reference therein for details). For this, we set two complex variables $A(t), A^*(t)$ that we define as 
\begin{align}
\psi(t) &= A(t) e^{i \Omega t} + A^*(t) e^{-i \Omega t} \\
\partial_t \psi(t) &= i \Omega A(t) e^{i \Omega t} -i \Omega A^*(t) e^{-i \Omega t} 
\label{eq:defA}
\end{align}
This transformation is invertible so that there is no hypothesis made at this stage. We are now looking for equations on $A$ and $A^*$. The first one is given by $\partial_t \psi = \partial \psi$ so that
\begin{equation}
\dot A e^{i \Omega t}  + \dot A^*  e^{-i \Omega t}  = 0 
\end{equation}
For the second equation, we will recast the equation of motion at first order in $\epsilon$ as 
\begin{equation}
\frac{d^2 \psi}{dt^2} + \Omega^2 \psi = - \gamma \epsilon \frac{d \psi}{dt} - \sigma \epsilon \psi - \epsilon \cos{(2 \Omega  t)} \psi + \epsilon \alpha \psi^3 = - \epsilon f(\psi, \dot \psi, t)
\end{equation}
with $\sigma = (1 - \Omega^2)/\epsilon \sim 1$. We can then find a second equation on $A$ and $A^*$ as
\begin{equation}
\partial_{tt} \psi = i \Omega \dot A e^{i \Omega t} - i \Omega \dot A^* e^{-i \Omega t} - \Omega^2 \psi
\end{equation}
so that one finally has
\begin{equation}
i  \Omega \dot A e^{i \Omega t} - i \Omega \dot A^* e^{-i \Omega t} = - \epsilon f(\psi, \dot \psi, t)
\end{equation}
The two equations can be summarized in the matrix form
\begin{equation}
\begin{bmatrix}
e^{i\Omega t} & e^{-i \Omega t} \\
i \Omega e^{i\Omega t} & -i \Omega e^{-i \Omega t}
\end{bmatrix}
\begin{bmatrix}
\dot A \\ \dot A^*
\end{bmatrix} =
\begin{bmatrix}
    0 \\  - \epsilon f
\end{bmatrix}
\end{equation}
and after inverting the matrix, one gets
\begin{equation}
    \dot A = \frac{i \epsilon}{2 \Omega} f(\psi, \dot \psi, t)e^{-i \Omega t}
    \label{eq:slowEnv}
\end{equation}

In equation (\ref{eq:slowEnv}), one has $\dot A \sim \epsilon$ so that $A$ varies slowly along time. In that case, one can average the right term of the previous equation over one fast oscillation at $\Omega_0$. After some computation, one finds
\begin{equation}
    \dot A= \epsilon \left( - \frac{\gamma}{2} + i \frac{\sigma}{2 \Omega} - i \frac{3 \alpha}{2 \Omega} |A|^2 \right) A + \frac{i \epsilon}{4 \Omega} A^* 
\end{equation}
If one assumes now that the unperturbed excitation is perfectly resonant, meaning $\omega = \omega_0$, one has $\Omega_0 =1$ and $\sigma=0$. In that case, one has
\begin{equation}
    \dot A= \epsilon \left( - \frac{\gamma}{2} - i \frac{3 \alpha}{2} |A|^2 \right) A + \frac{i \epsilon}{4} A^* 
    \label{eq:avg}
\end{equation}
If $A \ll 1$, the non-linear term can be neglected and the evolution equation can be expressed as
\begin{equation}
    \ddot A + \epsilon \gamma \dot A + \epsilon^2 \left[ \left(\frac{\gamma}{2} \right)^2 - \left(\frac{1}{4} \right)^2 \right] A = 0
\end{equation}
This admits solutions of the form
\begin{equation}
    A(t) = A_0 e^{(1 - 2\gamma)\epsilon t / 4} + B_0 e^{-(1 + 2\gamma)\epsilon t / 4}
\end{equation}
Thus we see that the solution exponentially grows if $\gamma < 1/2$ which corresponds to a minimal forcing of
\begin{equation}
    \epsilon > \epsilon_c = \frac{4}{\omega_0 \tau}
\end{equation}
As the amplitude will grow, one will not be able to neglect the non-linearity. Due to the latter, the system will reach a limit cycle such that $\dot A = 0$. The amplitude reached is fixed by the equation
\begin{equation}
     \left(\frac{\gamma}{2} + i \frac{3 \alpha}{2} |A|^2 \right) A = \frac{i A^*}{4} 
\end{equation}
that admits a complex solution $A_c \neq 0$ if $\epsilon > \epsilon_c$. Note that $-A_c$ is also solution, showing the degeneracy of the solution. The main point here is that $A$ goes toward one of the two fixed point $\pm A_c$. In particular, the phase of $A_c$ is fixed up to $\pi$.

\subsection{Adding a time defect}
We will now add a time defect in the excitation. For this, we define the detuning function $\phi_d(t) = 2 \omega \lambda(t) t$ with $\lambda(t)=\lambda$ for $0 < t < T_\lambda = \pi/\omega \lambda$ and $\lambda(t) = 0$ otherwise. The evolution equation now reads
\begin{equation}
\frac{d^2 \psi}{dt^2} + \frac{2}{\tau} \frac{d \psi}{dt} +  \omega_0^2(1 + \epsilon \cos{(2\omega  t + \phi_d(t))} ) (\psi - \alpha \psi^3 ) = 0
\end{equation}
Performing the same change of variable as before, we arrive at the evolution equation of the same form as Eq. \ref{eq:slowEnv} but that now contains the time defect.

We will now assume once again that the excitation and the pendulum are tuned so that $\Omega = 1$, and that $\phi(t)$ varies slowly compared to the natural oscillation of the pendulum. More precisely, we will assume that the detuning scales as $\lambda = \beta \epsilon \ll 1$ with $\beta \sim 1$. From this, one can get the averaged adimensionned equation as
\begin{equation}
    \dot A= \epsilon \left( - \frac{\gamma}{2} - i \frac{3 \alpha}{2} |A|^2 \right) A + \frac{i \epsilon}{4} A^* e^{2 i  \lambda(t) t}
    \label{eq:amplitudeTimeDefect}
\end{equation}

We now aim to find the value for $\lambda$ for which the  transition occurs. For this, we will place ourselves in the linear case, meaning $\alpha =0$. By defining a new amplitude $A(t) \rightarrow A(t) e^{-\epsilon \gamma t /2}$ the evolution equation now becomes
\begin{equation}
\dot A = \frac{i \epsilon}{4} A^* e^{2 i  \lambda(t) t}
\label{eq:AL}
\end{equation}
where $\lambda (t) = \lambda$ for $0 < t < T_\lambda$ (with adimensionned $T_\lambda = \pi/\lambda$) and is zero otherwise. We will now look how the amplitude is modified due to the perturbation. We start with solutions on each time domain
\begin{align}
A(t) &= A_0 e^{\Delta_0 t} + B_0 e^{-\Delta_0 /4}, \quad t < 0 \\
&= A_1 e^{(\Delta_\lambda + i \lambda) t} + B_1 e^{(-\Delta_\lambda +i \lambda ) t}, \quad 0 < t < T_\lambda \\
&= A_2 e^{\Delta_0 t} + B_2 e^{-\Delta_0 t}, \quad T_\lambda < t
\end{align}
where $\Delta_0=\epsilon/4$ and $\Delta_\lambda = \sqrt{(\epsilon/4)^2-\lambda^2}$ (that can be complex). The next thing we need is the continuity equations to go from one solution to the other.

Assuming that $A$ is continuous, $\dot A$ will also be continuous according to Eq. \ref{eq:AL} and the continuity of $e^{i \lambda(t) t}$ at $t=0$ and $t=T_\lambda$. From this, one can write the continuity equations at those time
\begin{equation*}
\begin{bmatrix}
1 & 1 \\
\Delta_\lambda + i \lambda   & -\Delta_\lambda + i \lambda  
\end{bmatrix}
\begin{bmatrix}
A_1 \\ B_1
\end{bmatrix} =
\begin{bmatrix}
1 & 1 \\
\Delta_0  & -\Delta_0
\end{bmatrix}
\begin{bmatrix}
A_0 \\ B_0
\end{bmatrix}
\end{equation*}

\begin{align*}
&\begin{bmatrix}
e^{(\Delta_\lambda + i \lambda) T_\lambda} & e^{(-\Delta_\lambda + i \lambda) T_\lambda} \\
(\Delta_\lambda + i \lambda) e^{(\Delta_\lambda + i \lambda) T_\lambda} & (-\Delta_\lambda + i \lambda)e^{-(\Delta_\lambda + i \lambda) T_\lambda}
\end{bmatrix}
\begin{bmatrix}
A_1 \\ B_1
\end{bmatrix} = \\
&\begin{bmatrix}
e^{\Delta_0 T_\lambda} & e^{-\Delta_0 T_\lambda} \\
\Delta_0 e^{\Delta_0 T_\lambda} & -\Delta_0 e^{-\Delta_0 T_\lambda}
\end{bmatrix}
\begin{bmatrix}
A_2 \\ B_2
\end{bmatrix}
\end{align*}
After some computation (and using that $e^{i \lambda T_\lambda} = -1$), one finds
\begin{align*}
&\begin{bmatrix}
A_2 \\ B_2
\end{bmatrix} = \\
&\begin{bmatrix}
- \alpha e^{-\Delta_0 T} \left( \frac{\Delta_0}{\Delta_\lambda } \sinh{(\Delta_\lambda T_\lambda)}  + \cosh{(\Delta_\lambda T_\lambda)} \right) & * \\
g(X) & * \\
\end{bmatrix}
\begin{bmatrix}
A_0 \\ B_0
\end{bmatrix}
\label{eq:transferMatrix}
\end{align*}
with $\alpha$ a positive prefactor that will be omitted and $g(X)$ a function that will be discussed later. We can now consider the cancellation of the left up term of the matrix. For this, we need to remember that $T_\lambda = \pi / \lambda$ so that we get
\begin{equation}
\Delta_\lambda T_\lambda = \pi \sqrt{\left( \frac{\epsilon}{4 \lambda} \right)^2 - 1}
\end{equation}
so that if one sets $X = \epsilon/4 \lambda$, the upper left coefficient reads up to positive constants
\begin{equation}
f(X) = -\frac{X}{\sqrt{X^2 - 1}} \sinh{\left( \pi \sqrt{X^2 - 1}\right)}  - \cosh{\left(\pi \sqrt{X^2 - 1} \right) }
\end{equation}
This function is of prime interest for the study of phase transition. Indeed, if the systems runs for a long time at negative times, the oscillator oscillates as $\Psi(t) = A_0 e^{\Delta_0 t} e^{i t} + h.c.$ as the exponentially decaying part can be neglected. Although the amplitude is not fixed, the oscillation phase is fully determined by the one of $A_0$. If one now waits long enough after the defect, the same remarks holds and the oscillation phase is determined by the one of $A_2$. Therefore, the sign of $f(X)$ directly encodes if there is a phase transition or not.

The graph of the previous function is shown in the main text. It is positive for $X < X_c \approx 0.68$. This means that the phase transition occurs for
\begin{equation}
\lambda < \lambda_c = \frac{\epsilon}{4 X_c} \epsilon \approx  0.37 \epsilon
\end{equation}
In other words, when the detuning is too large, no phase transition occurs while it occurs when the detuning is small. 

\subsection{Invariance of the equation}

As we have seen in the main text, the linear and non-linear transition diagrams are different. We aim here to analyse such differences by looking at the invariances of the evolution equation (\ref{eq:amplitudeTimeDefect}) in the linear and non-linear case under variation of external parameters. The later is written again here
\begin{equation}
    \dot A + \frac{1}{\tau} A=  - \frac{3 i \alpha \epsilon}{2} |A|^2 A + \frac{i \epsilon}{4} A^* e^{2 i  \lambda(t) t}
\end{equation}
Here the damping time $\tau$ is fixed and of the order of $\epsilon$. The four independant parameters of the equation are $\tau, \epsilon, \lambda, \alpha$ and we aim to find some invariance of the equation under some transformation on those parameters.

\subsubsection{Linear case}
The first step consists in time renormalisation. If we set the new time as $u = \epsilon t$, we get
\begin{equation}
    \partial_u A + \frac{1}{\epsilon \tau} A= \frac{i}{4} A^* e^{2 i  \Lambda(u) u}
\end{equation}
where the function $\Lambda(u)$ only depends on $\lambda/\epsilon$
\begin{equation}
    \Lambda (u) = \lambda/\epsilon, \qquad 0 \leq u \leq \pi \epsilon/|\lambda|
\end{equation} Next, we set as before $A \rightarrow A e^{-u/\epsilon \tau}$ that will not affect the phase of $A$. This finally gives the equation
\begin{equation}
    \partial_u A = \frac{i}{4} A^* e^{2 i  \Lambda(u) u}
\end{equation}
Therefore, we see here that the behavior of $A$ only depends on the ratio $\lambda/\epsilon$. In particular, this predicts automatically that transition curves must have equation of the form $\lambda_c = \alpha \epsilon$ which is consistent with the previous analysis.

One can also recover the anti-symmetry when $\lambda$ switches sign. If $\lambda \rightarrow - \lambda$, one gets
\begin{equation}
    \partial_u A = \frac{i}{4} A^* e^{-2 i  \Lambda(u) u}
\end{equation}
We can take the conjugation of this equation and set $A = i B^*$ and finally get
\begin{equation}
    \partial_u B = \frac{i}{4} B^* e^{2 i  \Lambda(u) u}
\end{equation}
and we recover the equation for positive values of $\lambda$. As we have performed complex conjugation to go from one to the other, this means that the winding index will also change sign, which is exactly what is observed.
\subsubsection{Non-linear case}
We now aim to see how the non-linearity affects-and breaks-the previous result. By performing $A \rightarrow A/\sqrt{\alpha}$, one can assume $\alpha = 1$. This means that the transition does not depend on the prefactor in front of the non-linear term but only on the type of non-linearity. Performing the same steps as before, we get an extra term in the evolution equation
\begin{equation}
    \partial_u A = - \frac{3i|A|^2 A}{2}  e^{-2u/\epsilon \tau} + \frac{i}{4} A^* e^{2 i  \Lambda(u) u}
\end{equation}
The exponential term due to damping breaks all the symmetries that where found from the linear case. The dependency with respect to $\epsilon$ breaks the $\lambda/\epsilon$ invariance. The anti-symmetry $\lambda \rightarrow - \lambda$ is also broken as the transformations previously performed will give the equation
\begin{equation}
    \partial_u B = \frac{3i|B|^2 B}{2}  e^{-2u/\epsilon \tau} + \frac{i}{4} B^* e^{2 i  \Lambda(u) u}
\end{equation}
and the sign in front of the non-linarity has changed. This is consistent with the fact that in presence of non-linearity, positive and negative detuning on a parametric oscillator are not equivalent but lead to sub- or supercrtical bifurcations.

\subsection{Analysis of the transition time}
We are now focusing on what happens near the transition meaning $X = X_c + \delta X$. In that case, one can approximate $f(X) \approx f'(X_c) \delta X$. On the other hand, one can show that the bottom left coefficient $g(X)$ previsouly introduced does not cancel at $X_c$ and takes value $g(X_c)$. We assume that before the defect, the system has run long enough so that one has $A(t<0) = A_0 e^{\Delta_0 t}$ so that $B_0 = 0$. The solution after the defect reads
\begin{align}
        A(t) &= A_2 e^{\Delta_0 t} + B_2 e^{-\Delta_0 t} \\
        &= f'(X_c) e^{\Delta_0 t} \delta X + g(X_c) e^{-\Delta_0 t}
\end{align}
As $\delta X \ll 1$, the dynamic of this function is first dominated by the second term that exponentially decays until both terms become of the same order. At this time, the exponential growth becomes the dominant term and $A(t)$ starts to grow again toward its new limit cycle. The time at which both terms are equal therefore gives a good hint on how the transition time $T_c$ should scale. From the previous equation, we see that this occurs for
\begin{equation}
    |\delta X| \sim e^{-2 \Delta_0 T_c}
\end{equation}
so that we get at the end (as $\delta X \sim \lambda - \lambda_c$)
\begin{equation}
    T_c \sim \ln |\lambda - \lambda_c|
\end{equation}
that is the scaling law discussed in the main text. For this argument to hold, the convergence time toward the limit cycle when the non-linearity has to be taken into account must be negligeable compared to $T_c$. This is always the case if one takes $\lambda-\lambda_c$ small enough. In this case, the system spends 'a lot of time' near zero in the phase space so that the non-linearity can be neglected for most of the convergence process.

In our numerical simulations, this time is computed the following way. We first compute $\psi(t)$ from a standard Runge-Kutta 4 method. We then compute the associated $A(t)$ through the transformation (\ref{eq:defA}). Running the simulation long enough ensures the convergence of $A(t)$ at large time toward its final value $A_c$. The transition time is then computed as the maximum time after the defect where $A(t)$ is still far enough from $A_c$. More precisely, one computes
 \begin{equation}
     T_c = \max \left\{ t>T_\lambda / |A(t)-A_c| > \epsilon/3 \right\}
 \end{equation}
 the choice of $\epsilon/3$ being a bit arbitrary but relevant for most of the considered cases. Changing the 1/3 factor changes the value of $T_c$ but not the observed scaling law $T_c \sim \ln |\lambda_c - \lambda|$

\subsection{Experimental methods and data analysis}
\begin{figure*}
    \centering
    \includegraphics[width=16cm]{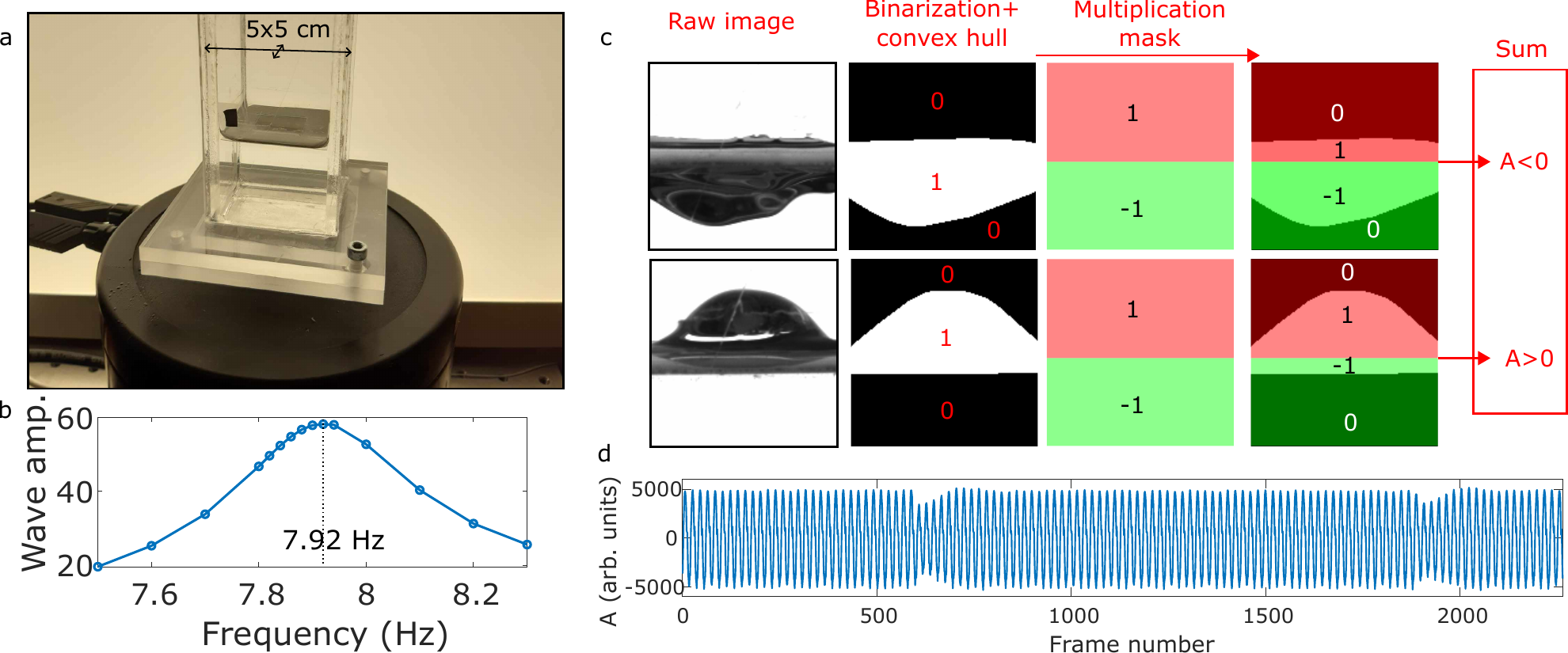}
    \caption{(a) Picture of the experimental setup: a tank filled with water is placed on a shaker. A plastic sheet is placed at the surface to prevent water waves to appear near the walls and a black mark is placed on the tank to recover its oscillatory motion.(b) Example of measured resonance with a peak at $7.06 \pm 0.01$ Hz.  (c) Image processing method: starting from raw image of waves, we perform binarization and convex hull computation. At this stage, the number of white pixels encodes the amplitude of the wave. In order to distinguish the case of waves above and below the plastic sheet, we multiply by an analyzing mask before summing the resulting image. (d) Example of wave amplitude measurement on a movie using the previous method.}
    \label{fig:2S}
\end{figure*}
\subsubsection{Experimental setup}
The experimental setup consists in a tank of 5x5 cm$^2$ filled with tap water as showm in Fig. \ref{fig:2S}a. The water depth is more than a couple of centimeters so that the exact value of the depth is irrelevant. The tank is placed on a shaker (B\& K 4809) and vibrated with a sinus signal generated with a generator (Rigol DG1022Z) and amplified with a dedicated amplifier (B\&K 2706). When the forcing amplitude is large enough, large amplitude Faraday waves oscillating at $f_0/2$ appear at the surface. In order to avoid friction on the walls, we place a transparent plastic sheet on top of the surface with a square hole of 2x2 cm$^2$ on its center. The sheet prevents the instability to occur near the walls and pin the meniscus of the fluid on the edges of the square. As the sheet is floating on the water surface, evaporation does not affect the relative position of the water surface with respect to the sheet, ensuring constant boundary conditions during the whole experiment (typically several hours). 

\subsubsection{Faraday waves and parametric oscillation}
For small deformation of the interface, the governing equation for the height field $h_k$ associated to wave vector $k$ at the surface of a vibrated fluid with acceleration $a \cos{(2 \omega t)}$ is \cite{douady_experimental_1990}
\begin{equation}
    \partial_{tt} h_k + \frac{1}{\tau} \partial_t h_k + (gk + \gamma k^3/\rho + ak \cos{(2 \omega t)}) h_k = 0
\end{equation}
where $\tau$ accounts for viscous dissipation and is measured experimentally to be $\tau = 0.9 \pm 0.03$ s, $g = 9.8$ m/s$^2$ is local gravity, $\gamma = 60$ mJ/m$^2$ is surface tension of water and $\rho = 1000$ kg/m$^3$ is water density. From this we can therefore estimate the adimensioned forcing as
\begin{equation}
    \epsilon = \frac{ak}{gk + \gamma k^3/\rho}
\end{equation}
We measure $k \approx 200$ m$^-1$ and our range of acceleration goes from $1.6$ m/s$^2$ to $3.6$ m/s$^2$, which corresponds to $0.13 < \epsilon < 0.30$.

\subsubsection{Resonance frequency of the interface}
The excitation frequency $f_0$ is chosen so that the forcing needed to trigger the instability at $f_0/2$. The frequency $f_0/2$ therefore corresponds to the frequency of a particular mode of the interface. We chose to excite the fundamental mode of the square hole that we measure experimentally by exciting different frequencies and measuring the wave amplitude in the cavity. A typical result is shown in Fig \ref{fig:2S}b. Depending on the exact experimental conditions (the room temperature appearing to be the most important), this frequency can drift a bit and typically changes between 7.84 Hz and 7.92 Hz depending on the day. For the measurement presented in the article in the linear case, we have measured at the beginning and the end of the experiment that resonance occurs at $7.92\pm 0.01$ Hz and therefore set $f_0 = 15.84$ Hz. For the non-linear case, we found resonance at $7.84\pm 0.01$ Hz and therefore picked $f_0 = 15.68$ Hz.

\subsubsection{Wave measurement and data analysis}

The wave amplitude is measured by placing a camera on the side of the tank. In order to have good temporal resolution, 60 images/second are taken, what ensures approximately 5 images per oscillation of the shaker and 10 images per oscillation of the wave.For small wave amplitude, we place a camera above the tank use a FCD technique described in \cite{wildeman_2017}. For larger amplitude, this technique eventually fails and we rather use a contrast analysis technique using a camera placed on the side. A LED panel is placed behind the tank and its position is adjusted so that the background appears white while the deformed liquid interface is black as shown in Fig. \ref{fig:2S}c. A black mark (see Fig. \ref{fig:2S}) is also placed on the tank and its position is tracked along time to recover the motion of the tank due to the shaking. This measurement is used to numerically compensate the motion of the tank so that it appears as non-moving in the wave movie. The algorithm used to recover the wave amplitude is described hereafter and summarized in Fig. \ref{fig:2S}d. We first apply a binarization process followed by a convex hull computation on raw images as shown in Fig. \ref{fig:2S}c. We then multiply the resulting image by a mask that discriminates whether the wave interface is above or below the plastic sheet, and sum all the pixels of the resulting image. The result of the sum is then with good approximation the wave amplitude. As we only focus on its phase, knowing the exact prefactor to go from our measurement to an amplitude measurement in physical units in nevertheless irrelevant in our case, and we therefore plot all the results in arbitrary units. A typical measurement result is shown in Fig. \ref{fig:2S}d.

In order to introduce a temporal defect as described in the main text, we use the frequency shift keying (FSK) functionnality of the function generator that allows to switch the excitation frequency between two values $f_0$ and $f_1$ depending on the value of an external $0/1$ trigger signal generated by the second channel of the generator. We then chose a given value of $\lambda$, set $f_1 = f_0(1+\lambda)$ and generate a trigger signal that is in the 1 state during $T_\lambda = 1/|f_0-f_1|$ and that is in the 0 state for $T_d = 8$ s. The time $T_d$ was chosen large enough so that after each defect of duration $T_\lambda$, the system could reach again a limit cycle with  properly defined phase. We then record the wave amplitude for approximately 30 seconds, so that the wave experiences at least three time a temporal defect. This allows to test the reproductibility of the result for each value of $\lambda$, which were not found to vary significantly from one realization to the other. The corresponding point in the phase diagram is then found by performing standard amplitude and phase measurement on the experimental results.
\begin{acknowledgments}
    The authors acknowledge Herve Lissek for the loan of the shaker and the Grant SNSF Eccellenza 181232 for financial support.
\end{acknowledgments}


\bibliography{apssamp}

\end{document}